# BEAM POSITION-PHASE MONITORS FOR SNS LINAC

S.S. Kurennoy, LANL, MS H824, Los Alamos, NM 87545, USA

*Abstract*

Electromagnetic modeling with MAFIA of the combined beam position-phase monitors (BPPMs) for the Spallation Neutron Source (SNS) linac has been performed. Time-domain 3-D simulations are used to compute the signal amplitudes and phases on the BPPM electrodes for a given processing frequency, 402.5 MHz or 805 MHz, as functions of the beam transverse position. Working with a summed signal from all the BPPM electrodes provides a good way to measure accurately the beam phase. While for an off-axis beam the signal phases on the individual electrodes can differ from those for a centered beam by a few degrees, the phase of the summed signal is found to be independent of the beam transverse position inside the device. Based on the analysis results, an optimal BPPM design with 4 one-end-shorted 60-degree electrodes has been chosen. It provides a good linearity and sufficient signal power for both position and phase measurements, while satisfying the linac geometrical constrains and mechanical requirements.

## 1 INTRODUCTION

Beam position-phase monitors in the SNS linac will deliver information about both the transverse position of the beam and the beam phase. Typical values for the beam position accuracy are on the order of 0.1 mm within 1/3 of the bore radius $r_b$ from the axis ($r_b$ is 12.5 mm to 20 mm for the normal conducting part of the linac). The BPPMs have a high signal processing frequency, equal to the microbunch repetition frequency in the linac, $f_b$=402.5 MHz (or its 2nd harmonics, 805 MHz). The beam phase measurement within a fraction of an RF degree is also required from the SNS linac BPPMs.

Various options for the transducers (pickups) of the SNS linac BPPMs have been studied using the EM code MAFIA [3] in [1,2]. Electrostatic 2-D computations are used to adjust the pickup cross-section parameters to form 50-Ω transmission lines. 3-D static and time-domain computations were applied to calculate the electrode coupling. Time-domain 3-D simulations with SNS beam microbunches passing through the BPPM at a varying offset from the axis were used to compute the induced voltages on the electrodes as functions of time. After that an FFT procedure extracted the amplitudes and phases of the signal harmonics at individual outputs, as well as the amplitude and phase of the combined (summed) signal, versus the beam transverse position. This information was used to choose an optimal BPPM design. Section 2 summarizes the results of this study.

In the SNS linac, there is a rare opportunity to put BPPMs and steering magnets inside the drift tubes in the drift-tube linac (DTL) to provide a better quality beam. This is due to the fact that every third drift tube (DT) is empty. The DTL RF fields, however, will produce an additional signal in BPPMs inside DTs at the DTL RF frequency 402.5 MHz that can exclude the BPPM signal processing at this frequency. For the coupled-cavity linac (CCL) there is no such problem, since its RF frequency is 805 MHz. In Sect. 3 we study the feasibility of using BPPMs in the DTL.

## 2 BPPM MODELING

### 2.1 BPPM Design

After considering a few possible pickup designs, we decided to choose a BPM design having 4 stripline electrodes with one end shorted. A MAFIA model of the BPM consists of a cylindrical enclosure (box) with 4 electrodes on a beam pipe, see Fig. 1. Each electrode covers a subtended angle of 60°. They are flush with the beam pipe, shorted at one end, and have 50-Ω connectors on the other end. For the CCL beam pipe radius $r_b$=20 mm, the electrode length along the beam is taken to be 40 mm. The 50-Ω electrode connectors are modeled by discrete elements, 50-Ω resistors in this case. This design is non-directional, provides a rigid mechanical structure, has all four connectors on one end, and therefore can be mounted close to quadrupoles or fit inside a DT.

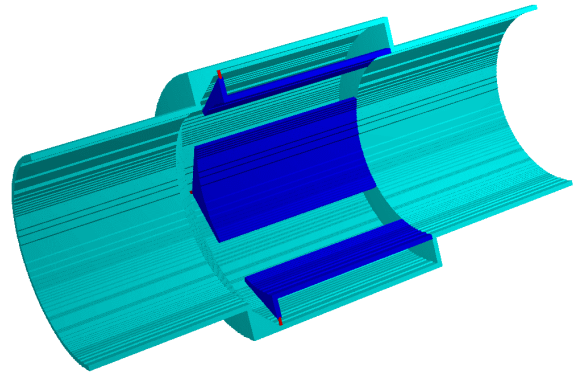

Figure 1: MAFIA model of BPPM (1/2-cutout) with cone-tapered box end and electrodes (dark) with ridged transitions to connectors (shown as red pins).

### 2.2 Position Measurements

Direct 3D time-domain computations with an ultra relativistic (β=1) bunch passing the structure at the axis or parallel to the axis have been performed. A Gaussian longitudinal charge distribution of the bunch with the total charge $Q$=0.14 nC and the rms length $\sigma$=5 mm, corresponding to the 56-mA current in the baseline SNS regime with 2-MW beam power at 60 Hz, was used in the simulations. Presently, the MAFIA time-domain code T3 cannot simulate the open (or waveguide) boundary conditions on the beam pipe ends for non-ultra relativistic ($\beta$<1) beams. The ultra relativistic MAFIA results are used to fix parameters of an analytical model of the BPPM at $\beta$=1, and then to extrapolate results for $\beta$<1 analytically.

To study the BPM linearity, we perform simulations with the beam bunch passing through the BPM with different transverse offsets. The amplitudes $\tilde{A}_P$ and the phases of the Fourier transforms of the induced voltages on all four ($P$=$R,T,L,B$ for the right, top, left and bottom) electrodes are calculated as the functions of the beam transverse position. The BPM position sensitivity was found to be equal to $20\log_{10}(\tilde{A}_R/\tilde{A}_L)/x \cong 1.4$ dB/mm. At high beam energies the signal power at 402.5 MHz changes between +4.6 dBm and –12.3 dBm for the beam position within a rather wide range, $\{x,y \in (-r_b/2, r_b/2)\}$, i.e. the signal dynamical range is 16.9 dB. The BPM linearity results are presented in Fig. 2. MAFIA data showing the horizontal signal log ratio $\ln(\tilde{A}_R/\tilde{A}_L)/2$ or the difference-over-sum $(\tilde{A}_R-\tilde{A}_L)/(\tilde{A}_R+\tilde{A}_L)$ for different vertical beam positions overlap, so that it is difficult to distinguish between the five interpolating lines in each group. We can conclude that this BPM design is insensitive to the beam position in the direction orthogonal to the measured one, and has a good linearity.

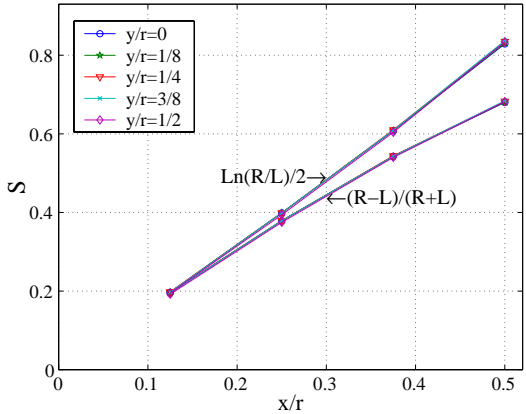

Figure 2: Signal ratio $S$ at 402.5 MHz versus the beam horizontal displacement $x/r_b$, for a few values of the beam vertical displacement $y/r_b$.

### 2.3 Analytical Model of BPM

Assuming an axial symmetry of the beam pipe, the signals on the BPM electrodes of inner radius $r_b$ and angle $\varphi$ can be calculated by integrating induced currents within the electrode angular extent. For a pencil beam bunch passing the BPM at the transverse position $x=r\cos\theta$, $y=r\sin\theta$ at velocity $v=\beta c$, the signals are (e.g., [4]):

$$E(f,r,\theta) = C \frac{\varphi}{2\pi} \left[ \frac{I_0(gr)}{I_0(gr_b)} + \frac{4}{\varphi} \sum_{m=1}^{\infty} \frac{I_m(gr)}{I_m(gr_b)} \sin\left(m\left(\frac{\phi}{2}+\mu\right)\right) \cos(m(\theta-\nu)) \right] \quad (1)$$

where $E$=$R,T,L,B$ are the Fourier amplitudes at frequency $f$ of the voltages on the electrodes, $(\mu,\nu)$ are $(0,0)$ for $R$, $(0,\pi/2)$ for $T$, $(\pi,0)$ for $L$, $(\pi,\pi/2)$ for $B$, and $I_m(z)$ are the modified Bessel functions. All dependence on frequency and energy is through $g=2\pi f/(\beta\gamma c)$, and overall coefficient C depends on the beam current.

The parameters $r_b$ and $\varphi$ can be considered as "free" parameters of the model. To find their effective values, we fit with Eqs. (1) the MAFIA results for $\beta$=1 at 402.5 MHz for the ratio $S/(x/r_b)$, where $S$ is either $\ln(\tilde{A}_R/\tilde{A}_L)/2$ or $(\tilde{A}_R-\tilde{A}_L)/(\tilde{A}_R+\tilde{A}_L)$. The best fit to the numerical data was obtained [1] with the effective parameters $r_{eff}$=$1.17 r_b$, $\varphi_{eff}$=$1.24\varphi$ (=74.5°), where $r_b$=20 mm, $\varphi$=60° are the geometrical values. Matching the amplitude of 402.5-MHz harmonics from an on-axis relativistic SNS beam bunch with Eqs. (1) fixes the constant C=1.232 V. Then the model reproduces MAFIA-computed 402.5-MHz signal amplitudes for the displaced beams with accuracy 1-2%. Assuming the effective parameters of the model applicable also at lower beam velocities, we extrapolate $\beta$=1 results to $\beta$<1. The signal power level for the on-axis beam is reduced by about 9 dB at $\beta$=0.073 (2.5 MeV). For the strongest signal in the beam displacement range $(-r_b/2, r_b/2)$ both vertically and horizontally, this reduction is 4.4 dB, and for the weakest one it is 12.9 dB. As a result, the dynamical range of the 402.5-MHz signal would increase from about 17 dB for $\beta$=1 to about 25 dB at $\beta$=0.073, if the same radius of BPM were assumed.

### 2.4 Phase Measurements

Two candidates for the beam phase detectors for the SNS linac – the capacitive probes and BPMs, either with signals from individual electrodes or with summed signals – have been studied and compared in Ref. [2]. MAFIA simulations with an ultra-relativistic beam, as well as measurements (for the capacitive probes), have shown a strong dependence of the measured beam phase on the transverse beam position inside a probe, when signals are picked up from individual connectors.

For an off-axis beam, the signal phases from individual electrodes can differ from those for a centered beam by a few degrees, while the phase of a summed signal remains the same within the computation errors (0.1-0.2°), even for the beam offsets as large as the pipe half-aperture. It is illustrated by Fig. 3 (the error bars are shown only for the summed signal). In the capacitive probe, the phase deviations from the centered beam phase grow as the beam offset increases, approaching 1 degree difference for large (half-aperture) offsets at the frequency 402.5 MHz. Based on the results of the analysis [2], we have chosen the BPMs with summed signals from all electrodes as the beam phase detectors in the SNS linac.

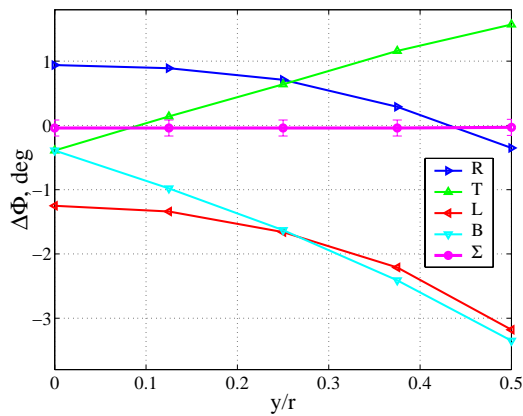

Figure 3: 402.5-MHz signal phases on BPM electrodes and for summed signal versus beam vertical displacement $y/r_b$, for the beam horizontal offset $x/r_b=1/4$.

## 3 BPPM IN DTL

We consider the tightest spot, the third DT in the 2nd DTL tank. The DT length along the beam is about 8 cm and its beam-pipe inner radius is 12.5 mm. The pickup design with four 60° electrodes is similar to that in the CCL, we only reduce the transverse dimensions and take the electrode length to be 32 mm. The beam-induced signals at the pickup electrodes are computed using MAFIA simulations with an ultra relativistic beam. The Fourier harmonics amplitudes for the on-axis beam are $\tilde{A}_1$=0.190 V at 402.5 MHz and $\tilde{A}_2$=0.356 V at 805 MHz. We extrapolate these $\beta=1$ results to the H$^-$-beam energy of 7.5 MeV ($\beta_1$=0.126) analytically as described in Sect. 2.3. For the first harmonics, the ratio $S(\beta_1)/S(1)$=0.80 results in the beam-induced signal amplitude 0.152 V at 7.5 MeV. For 805 MHz, the ratio $S(\beta_1)/S(1)$=0.455 gives the signal amplitude 0.162 V. We want to compare these numbers with the signal amplitudes induced by the RF field in the DTL BPPM.

To calculate the signal power on the BPPM electrodes induced by the 402.5-MHz RF field in the DTL tank, we put the DT with the BPPM inside in a cylindrical pillbox having the length of 96 mm (twice that of DTL half-cell), and adjust the pill-pox radius to tune the frequency of its lowest axisymmetric mode to 402.5 MHz. Integrating the electric field of the computed eigenmode along the electrode connector gives $V_{con}$, and along the beam axis $V_{ax}$. We calculate the scaling factor as the ratio of the on-axis voltage given by SUPERFISH design computations, $V_{ax-SF} = 2.96 \cdot 10^5$ V, to $V_{ax}$. Multiplying $V_{con}$ by this scaling factor gives the RF-induced voltage amplitude $V_{ind}$. The results are listed in Table 1 for a few different pickup positions inside the DT. Here $z_c$ is the longitudinal coordinate of the BPPM electrode center relative to the DT middle point, and $z_g$ is the same for the BPM annular gap center. Since the electrode length is 32 mm and the gap is 4 mm wide, we have $z_g - z_c = 32/2 + 4/2 = 18$ mm.

Table 1: RF-induced signals versus BPPM position.

| $z_c$, mm | $z_g$, mm | $V_{ind}$, V | $P$, dBm |
|---|---|---|---|
| 8 | 24 | 15.30 | 33.69 |
| 0* | 18 | 3.28 | 20.32 |
| -8 | 10 | 0.72 | 7.15 |
| -12 | 6 | 0.36 | 1.13 |
| -16 | 2 | 0.23 | -2.77 |
| -18 | 0** | 0.22 | -3.15 |
| -20 | -2 | 0.24 | -2.40 |

\* The electrode center is at the DT center
\*\* The gap center is at the DT center

Obviously, placing the BPPM gap near the DT center reduces the RF-induced signal significantly. This is due to the axial symmetry of the RF field, which penetrates effectively only through the annular gap, but not through the longitudinal slots between the electrodes. For the optimal BPM position inside the DT, the RF-induced voltages have the same order of magnitude as the beam-induced ones: 0.22 V versus 0.15 V at 402.5 MHz, and versus 0.16 V at 805 MHz. While this prevents us from processing BPPM signals at 402.5 MHz, we can be sure that this BPPM inside the DT can operate with the RF power on without damage to the cables or electronics, and the filtering out the 402.5-MHz signal will present no problem for the BPM signal processing at 805 MHz.

## 4 SUMMARY

Electromagnetic MAFIA modeling of the SNS linac BPPMs has been performed. Based on the analysis results, an optimal pickup design with 4 one-end-shorted 60-degree electrodes has been chosen. It provides a good linearity and sufficient signal power for both position and phase measurements, while satisfying the geometrical and mechanical requirements, see in [1,2]. The feasibility of using BPPMs in the SNS drift-tube linac is demonstrated.

The author acknowledges useful discussions with J.H. Billen, J.F. O'Hara, J.F. Power, and R.E. Shafer.

## REFERENCES

[1] S.S. Kurennoy, "Electromagnetic Modeling of Beam Position Monitors for SNS Linac", Proc. EPAC 2000; http://accelconf.web.cern.ch/accelconf/e00/PAPERS/WEP2A08.pdf
[2] S.S. Kurennoy, "Beam Phase Detectors for Spallation Neutron Source Linac", Proceed. EPAC 2000; http://accelconf.web.cern.ch/accelconf/e00/PAPERS/WEP2A07.pdf
[3] MAFIA Release 4.20, CST GmbH, Darmstadt, 1999.
[4] R.E. Shafer, in AIP Conf. Proc. 319, 1994, p. 303.
[5] S.S. Kurennoy, "BPMs for DTL in SNS Linac", Tech memo SNS:00-54, Los Alamos, 2000.